\documentclass{article}

\usepackage[utf8]{inputenc}
\usepackage[british]{babel}
\usepackage{csquotes}

\usepackage{amsmath}
\usepackage{amssymb}
\usepackage{authblk}
\usepackage{mathrsfs}

\usepackage{outlines}

\usepackage{caption}
\usepackage{subcaption}
\usepackage{graphicx}
\usepackage{overpic}

\usepackage{orcidlink}

\usepackage{float}
\usepackage[final]{showlabels}

\usepackage{accents}
\usepackage{constraints}

\newcommand{\mytitle}{Cosmological initial data without periodic boundary conditions}

\usepackage{hyperref}
\hypersetup
{
    colorlinks=true,
    linktocpage=true,
    linkcolor={red},
    anchorcolor={black},
    citecolor={green},
    menucolor={red},
    unicode=true,
    pdftitle={\mytitle},
    pdfauthor={Károly Z. Csukás}
}

\newcounter{mnotecount}

\newcommand{\mnotex}[1]
{
  \protect{\stepcounter{mnotecount}}
    $^{\mbox{\footnotesize $\bullet$\themnotecount}}$
    \marginpar{\color{red}\raggedright\tiny\em
  $\!\!\!\!\!\!\,\bullet$\themnotecount: #1}
}

\usepackage[backend=biber,style=numeric-comp,sorting=none,backref=true]{biblatex}
\title{\mytitle}
\author{Károly Csukás\,\orcidlink{0000-0002-2408-1103} \footnote{E-mail address:{\tt csukas.karoly@wianer.hu}}}

\affil{Wigner RCP, H-1121 Budapest, Konkoly Thege Mikl\'{o}s \'{u}t 29--33, Hungary}

\begin{document}
  \maketitle

  \begin{abstract}
    We apply the parabolic-hyperbolic formulation of the Einstein constraint equations to generate cosmological initial data. The freely specifiable geometric data correspond to flat Friedmann--Lema\^itre--Robertson--Walker background, while the matter sector contains localized anisotropic perturbations of a perfect fluid. Unlike the standard approach, our method evolves the constraints outward from regular data at the origin and therefore requires no boundary conditions. This makes our method well suited as a starting point in investigating systematic biases introduced by commonly adopted boundary conditions, such as periodic boundaries. A second advantage concerns uniqueness: standard elliptic solvers may fail when multiple solutions exist, whereas solving the constraints as a well-posed evolutionary system always yields a unique solution. To demonstrate our method, we implement it numerically and generate cosmological initial data with localized anisotropic perfect fluid perturbations.
  \end{abstract}

  \newpage

  \tableofcontents

  \newpage

  \section{Introduction}

  One of the most impressive features of general relativity is its ability to describe the behavior of our whole Universe. We can get surprisingly far by considering linear perturbations over the homogeneous isotropic Friedmann--Lema\^itre--Robertson--Walker (FLRW) spacetime. However, there are inherently nonlinear phenomena in cosmology, such as formation of primordial black holes or homogenizing mechanisms in the early Universe. Such phenomena are best studied using the methods of numerical relativity \cite{Aurrekoetxea:2024ypv}. The most common approach is to consider the $4$-dimensional spacetime emerging as a $1$-parameter family of $\Sigma_t$ $3$-dimensional time slices resulting from the evolution of some initial data specified on one of these hypersurfaces, $\Sigma_0$. Whenever we consider such a $3+1$ decomposition of the problem the Einstein equation splits into a set of evolution equations and a set of constraint equations. The latter must be solved already on the initial $\Sigma_0$ slice. This paper introduces a new method to produce constraint satisfying cosmological initial data leveraging the parabolic-hyperbolic form of the constraints \cite{Racz:2014sua, Racz:2014gea, Racz:2015mfa}.

  We usually solve the Einstein constraint equations using one of the methods based on the works of Lichnerowicz \cite{lichnerowiczLintegrationEquationsGravitation1944} and York \cite{York:1973ia}. All these methods share a common feature, namely they are based on a conformal decomposition of the gravitational degrees of freedom and they result in a nonlinear system of elliptic partial differential equations. As such, we have to prescribe boundary conditions on the boundaries of the computational domain. In astrophysical configurations, where we want to model isolated self-gravitating systems, asymptotic flatness gives us physically well-motivated outer boundary conditions enforcing the necessary falloff rates of the fields. However, in cosmology it is less obvious what would be the right approach.

  The most popular choice is to enforce periodic boundary conditions on the faces of a cubical computational domain \cite{Aurrekoetxea:2024ypv}. In particular, we may argue that we want to simulate some representative patch of an infinite Universe. Since the Universe is believed to be isotropic and homogeneous on sufficiently large scales, periodic boundary conditions would be a good approximation to the large-scale Universe. However, the systematic bias associated with this choice of boundary condition is extensively debated in the literature. The most obvious effect is that this boundary condition enforces $T^3$ toroidal spatial topology, which puts a constraint on the spatial curvature. In particular, nonzero curvature is only possible if there is a region with negative curvature \cite{kazdanScalarCurvatureConformal1975}, which is nicely demonstrated in \cite{Garfinkle:2023vzf}, where the authors took advantage of this property. However, in general this constraint is quite restrictive, for example, it only allows the spatially flat solution among the FLRW spacetimes. Another effect of using periodic boundary conditions is the introduction of an artificial discrete symmetry with a gauge dependent scale. Ref \cite{Racz:2020nsv} demonstrated that the structures formed in the simulations reflect this symmetry and gives estimates how large the spatial extent of the simulation should be to keep this effect under a certain threshold. The problem with periodic boundary condition is even more severe: even if we use an evolution scheme that avoids using boundary conditions, the periodicity of initial data is already sufficient to significantly suppress curvature growth during evolution \cite{Coley:2023iol}. Although periodic boundary conditions are the most prevalent in the literature, there are other approaches. For example, \cite{Niemeyer:1999ak,Corman:2022rqo} matched to an FLRW solution on the outer boundary, while \cite{East:2016anr} used de Sitter for the same purpose. Furthermore, \cite{Brady:2026evi} considers asymptotically flat spacetimes specifically to study the bias introduced by periodicity.

  Here, we introduce a method that solves the constraints without enforcing any boundary conditions. To do so we solve the parabolic-hyperbolic form of the constraints \cite{Racz:2014sua, Racz:2014gea, Racz:2015mfa}. So far the only study using this formalism in the cosmological context is \cite{estrada-llestaNumericalStabilityHyperbolic2026}, where the authors considered $\Lambda$-vacuum on $T^3$ topology. Here we consider localized, anisotropic perturbations in a perfect fluid matter sector. Using a spherical foliation of the initial data surface we impose regularity conditions at the origin and integrate the constraints in the increasing radial direction to an arbitrary radius.

  Besides eliminating the need for boundary conditions, our method has another useful feature. It is known that the elliptic formulation does not always have unique solutions. Uniqueness of solutions to the momentum constraint breaks down specifically when we consider cosmological setup, because the conformal FLRW metric accommodates Killing vectors \cite{Garfinkle:2020iup}. Even if we consider moments of time symmetry in the initial data, when the momentum constraint becomes trivial we still have to solve the Hamiltonian constraint. The toy models of \cite{Baumgarte:2006ug,Baumgarte:2025vvs} show that the solution of the Hamiltonian constraint is, in general, non-unique neither in the astrophysical, nor in the cosmological setup. One way around this problem is to use conformally rescaled matter fields, or to interpret the Hamiltonian constraint as an algebraic equation as done in \cite{Aurrekoetxea:2022mpw}. The issue with non uniqueness of the solution is that standard methods for solving elliptic equations find the solutions iteratively. If there are multiple solutions the iteration might start jumping between the solutions instead of converging toward one. Furthermore, \cite{Baumgarte:2025vvs} shows that one branch of their solution is optimal for studying primordial black hole formation, while the other branch is better suited to study inflation. Therefore, it is not enough to ensure we find a solution, we need control over which branch we find. Indeed, \cite{Garfinkle:2020iup} devised a clever algorithm for solving the momentum constraint taking into account the nature of the solution, but the general problem remains unresolved. In contrast, the parabolic-hyperbolic formulation always results in a unique solution which can be reliably generated by any of the known evolutionary methods.

  The first half of the paper, Section \ref{sec:methods}, introduces the setup. Specifically Section \ref{sec:para-hyper} reviews the parabolic-hyperbolic formulation of the Einstein constraint equations. Section \ref{sec:flrw} specifies the cosmological setting, then Section \ref{sec:orig} introduces the regularity conditions at the origin. The second half, Section \ref{sec:num}, presents a numerical solution. Finally, we close with the discussion in Section \ref{sec:discussion}.

  Throughout this paper we use units such that $c=8\pi G=1$, and the signature of the ambient spacetime is assumed to be $(-,+,+,+)$. Regarding hypersurface geometry we follow the sign conventions of Wald \cite{waldGeneralRelativity1984} for the extrinsic curvature and acceleration vector, i.e. $K_{ab}=\boldsymbol{+}\tfrac{1}{2}\pounds_nh_{ab}$, $\dot{n}_a=\boldsymbol{+}n^bD_bn_a=\boldsymbol{-}D_aN/N$.

  \section{Geometric setup}
  \label{sec:methods}

  In general relativity the part of the initial data representing the geometrical content consists of the triplet $(\Sigma,h_{ab},K_{ab})$, where $\Sigma$ is a $3$-dimensional manifold, while $h_{ab}$ and $K_{ab}$ are symmetric tensor fields on $\Sigma$. Once $\Sigma$ is embedded into a $4$-dimensional spacetime, $h_{ab}$ is interpreted as the metric induced on $\Sigma$, and $K_{ab}$ as its extrinsic curvature. The initial data, however, cannot be arbitrary, as it has to satisfy the constraint equations \cite{waldGeneralRelativity1984}
  \begin{align}
   \Rthree+K^2-K_{ab}K^{ab}&=2\ee,\label{eq:Ham3p1}\\
   D_bK^b{}_a-D_aK&=-S_a,\label{eq:mom3p1}
  \end{align}
  where $D_a$ is the covariant derivative compatible with $h_{ab}$, $\Rthree$ is the corresponding Ricci scalar, $K=h^{ab}K_{ab}$ is the trace of extrinsic curvature. The source terms, $\ee$ and $S_a$, are related to the stress-energy tensor of the matter fields as $\ee=n^an^bT_{ab}$ and $S_a=-h_a{}^bn^cT_{bc}$, with $n_a$ being the future directed unit normal $1$-form of $\Sigma$.

  In this paper we consider matter content that satisfy the following two conditions. First, the matter must be unconstrained on the initial data hypersurface $\Sigma$. Second, the matter sources $\ee$ and $S_a$ must be independent of metric and extrinsic curvature degrees of freedom. The goal of these conditions is to keep the discussion as general as possible. Without any of these conditions we would be forced to consider a specific matter model that provides us with a specific matter constraint equation, or a specific formula telling us how the stress-energy components depend on the geometry.

  To see that these two conditions are not overly restrictive, we could consider a perfect fluid matter sector with stress-energy tensor
  \begin{equation}
   T_{ab}=\rho u_au_b+P(g_{ab}+u_au_b),\label{eq:Tfluid}
  \end{equation}
  where $\rho$ is the energy density, $P$ is the pressure in the comoving frame, and $u^a$ is the $4$-velocity of the fluid \cite{waldGeneralRelativity1984}. Then the source terms appearing in Eqns \eqref{eq:Ham3p1}-\eqref{eq:mom3p1} are the energy density $\ee=(\rho+P) w^2-P$, and the energy-current density $S_a=w(\rho+P)(h_a{}^bu_b)$ measured by the observer moving along the timelike unit vector field $n^a$, and $w=-n_au^a$ is the Lorentz factor. 
  The matter equations resulting from $\nabla^aT_{ab}=0$ yield evolution equations for $\ee$ and $S_a$. Importantly, the matter content is unconstrained and we are free to specify $\ee$ and $S_a$ on $\Sigma$, and only then solve the constraint equations \eqref{eq:Ham3p1}-\eqref{eq:mom3p1}.

  For the sake of brevity we will refer to perfect fluid as the matter content with the understanding, that our argument applies to a more generic class of matter models. For exactly this same reason we will not discuss the interpretation of the sources $\ee$ and $S_a$, since this interpretation necessarily depends on the specific model.

  \subsection{The parabolic-hyperbolic form of the constraints}
  \label{sec:para-hyper}

  The formalism presented in \cite{Racz:2014sua,Racz:2014gea,Racz:2015mfa} relies on the assumption that $\Sigma$ can be foliated by a $1$-parameter family of $\mathcal{S}_r$ level sets of some function $r:\Sigma\rightarrow\mathbb{R}$. Then we can construct the unit normal $1$-form $\nhat_a$ and the $\gammahat_{ab}$ metric induced on the $\mathcal{S}_r$ surfaces. We also choose an evolution vector field $r^a$, normalized as $r^aD_ar=1$, whose integral curves intersect each $\mathcal{S}_r$ surface exactly once. Then we can decompose the fundamental variables into quantities intrinsic to the foliating surfaces. The (secondary) lapse $\Nhat=\nhat_ar^a$, (secondary) shift $\Nhat^a=\gammahat^a{}_br^b$, and the $\gammahat_{ab}$ induced metric represent the degrees of freedom we have in $h_{ab}$. Similarly the projections $\kkappa=\nhat^a\nhat^bK_{ab}$, $\kk_a=\gammahat_a{}^b\nhat^cK_{bc}$, $\KK=\gammahat^{ab}K_{ab}$, and $\Kc_{ab}=\gammahat_a{}^c\gammahat_b{}^dK_{cd}-\frac{1}{2}\gammahat_{ab}\KK$ represent all the degrees of freedom we have in the extrinsic curvature. We also introduce the normal and tangential pressure $\pph=\nhat_aS^a$ and $\pp_a=\gammahat_a{}^bS_b$.

  So far we have not put any restrictions on the topology of the $\mathcal{S}_r$ surfaces. Indeed, there are working examples with planar foliation \cite{Racz:2017krc}, but the whole argument of this paper relies on the behavior of the constraints when we use topological $2$-spheres. Following \cite{Racz:2017krc}, on a single $\mathcal{S}_{r_0}$ leaf of the foliation we introduce $(\vartheta,\varphi)$ standard spherical coordinates and a complex null dyad $q_a$ normalized as $q_aq^a=0$, $q_a\qbar^a=2$, and $q_{ab}=\frac{1}{2}(q_{a}\qbar_{b}+q_{b}\qbar_{a})$, where $q_{ab}$ is a reference unit sphere metric with respect to the $(\vartheta,\varphi)$ coordinates. We extend these quantities to the whole $\Sigma$ manifold by Lie-dragging them along $r^a$. Finally we take the dyad components of the vector and tensor quantities introduced in the previous paragraph as $\NN=q^a\Nhat_a$, $\aa=\frac{1}{2}\gammahat_{ab}q^a\qbar^b$, $\bb=\frac{1}{2}\gammahat_{ab}q^aq^b$, $\kk=q^a\kk_a$, $\Kcqq=\Kc_{ab}q^aq^b$, $\pp=q^a\pp_a$. Angular derivatives of these spin-weighted quantities will be expressed as the Newman--Penrose $\eth$ and $\ethb$ operators. This procedure replaces $(\Sigma,h_{ab},K_{ab},\ee,S_a)$ with the algebraically equivalent set $(\mathcal{S}_r,\Nhat,\NN,\aa,\bb,\kkappa,\kk,\KK,\Kcqq,\ee,\pph,\pp)$ of spin-weighted variables.

  The parabolic-hyperbolic form of the constraints \eqref{eq:Ham3p1}-\eqref{eq:mom3p1} then read as \cite{Racz:2015mfa,Racz:2017krc}
  \begin{multline}
      \label{eq:phN}
      \Kstar\left[\partial_\rho\Nhat-\tfrac12\,\NNt\,\ethb\Nhat-\tfrac12\,\NNtbar\,\eth\Nhat\right]\\
      -\tfrac12\,\dd^{-1}\Nhat^2\left[\,\aa\left\{\eth\ethb\Nhat-\BB\,\ethb\Nhat\right\}-
      \bb\left\{\ethb^2\!\Nhat-\tfrac12\,\AAbar\,\ethb\Nhat-\tfrac12\,\CCbar\,\eth\Nhat\right\}+``cc"\right]\\
      -\Ascr\,\Nhat-\Bscr\,\Nhat^{\,3}=\ee\Nhat^3\,,
  \end{multline}
  \begin{equation}
      \label{eq:phk}
      \partial_\rho\kk-\tfrac12\,\NNt\,\ethb\kk-\tfrac12\,\NNtbar\,\eth\kk-\tfrac12\,\Nhat\,\eth\KK+\ff=-\Nhat\pp\,,
  \end{equation}
  \begin{equation}
      \label{eq:phK}
      \partial_\rho\KK-\tfrac12\,\NNt\,\ethb\KK-\tfrac12\,\NNtbar\,\eth\KK-
      \tfrac{1}{2}\,\Nhat\,\dd^{-1}\Big\{\,\aa(\eth\kkbar+\ethb\kk)-\bb\,\ethb\kkbar-\bbbar\,\eth\kk\,\Big\}+\FF=\Nhat\pph\,,
  \end{equation}
  where the coefficients $\Ascr$, $\Bscr$, and the source terms $\ff$, $\FF$, in \eqref{eq:phN}, \eqref{eq:phk}, and \eqref{eq:phK}, are given as
  \begin{equation}\label{eq:phA}
  	\Ascr=\partial_r\Kstar-\tfrac12\,\NNt\,\ethb\Kstar-\tfrac12\NNtbar\,\eth\Kstar+\tfrac12\,\Big[\Kstar{}^2+\Kstar_{ab}\Kstar{}^{ab}\Big]\,,
  \end{equation}
  \begin{equation}\label{eq:phB}
  	\Bscr=-\tfrac12\,\Big[\Rhat+2\,\kkappa\,\KK+\tfrac12\,\KK^2-\dd^{-1}[2\,\aa\,\kk\,\kkbar-\bb\,\kkbar^2-\bbbar\,\kk^2]-\Kc{}_{ab}\Kc{}^{ab}\Big]\,,
  \end{equation}
  \begin{multline}\label{eq:phf}
    \ff=-\tfrac12\,\Big[\kk\,\eth\NNtbar+\kkbar\,\eth\NNt\Big]-\left[\kkappa-\tfrac12\,\KK\right]\eth\Nhat+\Kstar\,\kk-\Nhat\Big[\eth\kkappa+q^a\dot{\nhat}{}^b\Kc{}_{ab}-q^a\widehat{D}^b\Kc{}_{ab}\Big]\,,
  \end{multline}
  \begin{align}\label{eq:phF}
    \FF=\tfrac{1}{4}\Nhat\,\dd^{-1}\Big\{2\,\aa\,\BB\,\kkbar-\bb(\,\CCbar\,\kk+\AAbar\,\kkbar)+``cc"\Big\}
    & -\dd^{-1}\Big[(\aa\,\kkbar-\bbbar\,\kk)\,\eth\Nhat+``cc"\Big]\nonumber \\ & +\Big[\Kc_{ab}{\Kstar}{}^{ab}-\left(\kkappa-\tfrac12\,\KK\right)\Kstar\Big]\,
  \end{align}
  with $``cc"$ denoting the complex conjugate of the preceding terms, while the explicit form of the terms, such as $\Kstar_{ab}{\Kstar}{}^{ab}$, $\Kc_{ab}{\Kstar}{}^{ab}$, $\Kc_{ab}{\Kc}{}^{ab}$, $q^a\dot{\nhat}{}^b\Kc{}_{ab}$, $q^a\widehat{D}^b\Kc{}_{ab}$, can be found in \cite{Racz:2017krc}.

  The momentum constraint \eqref{eq:phk}-\eqref{eq:phK} comprise a symmetric hyperbolic system for $\kk$ and $\KK$, while the Hamiltonian constraint \eqref{eq:phN} is a parabolic partial differential equation for $\Nhat$. The direction in which the latter (and therefore the whole system) can be integrated is determined by the factor $\Kstar=\frac{1}{2}\gammahat^{ab}\pounds_r\gammahat_{ab}-\widehat{D}_a\Nhat^a$, where $\pounds_r$ denotes the Lie derivative with respect to the $r^a$ vector field. Considering a foliation by topological $2$-spheres means that we specify the free data $(\NN,\aa,\bb,\kkappa,\Kcqq,\ee,\pph,\pp)$ on $\Sigma$ along with Cauchy data $(\Nhat|_{r_0},\kk|_{r_0},\KK|_{r_0})$ on some $\mathcal{S}_{r_0}$ surface, then Eqns \eqref{eq:phN}-\eqref{eq:phK} determine the solution for $r>r_0$.

  This paper exploits this behavior by choosing $r_0$ to be the origin, then we specify Cauchy data $(\Nhat|_{r_0},\kk|_{r_0},\KK|_{r_0})$ compatible with regularity conditions at the origin, and then integrate the constraints \eqref{eq:phN}-\eqref{eq:phK} outward up to an arbitrary, user-specified radius. This way we replace the often not well-motivated boundary conditions used by elliptic methods with well-motivated regularity conditions at the origin.

  \subsection{Cosmological initial data}
  \label{sec:flrw}

  Our strategy to construct cosmological initial data is to set the free data $(\NN,\aa,\bb,\allowbreak\kkappa,\Kcqq)$ according to some slicing of the spatially flat FLRW solution, while introducing localized anisotropic perturbations in the $(\ee,\pph,\pp)$ matter variables.

  The values of the non-vanishing spin-weighted variables computed from a $t=const$, $r=const$ slicing of the
  \begin{equation}
    \mathrm{d}s^2=-\mathrm{d}t^2+a^2(t)[\mathrm{d}r^2+r^2(\mathrm{d}\vartheta^2+\sin^2\vartheta\mathrm{d}\varphi^2)],
  \end{equation}
  spatially flat FLRW metric \cite{ellisRelativisticCosmology2012} are
  \begin{align}
    \Nhat&=a(t),&\aa&=a^2(t)r^2,\label{eq:FLRWspin1}\\
    \kkappa&=\frac{\dot{a}(t)}{a(t)},&\KK&=2\frac{\dot{a}(t)}{a(t)},\label{eq:FLRWspin2}
  \end{align}
  and it follows that $\Kstar=2/r$. The Hamiltonian constraint also enforces the relation $\kkappa=\sqrt{\ee_0/3}$, where the energy $\ee=\ee_0+\delta\ee$ is split into a homogeneous background value and an anisotropic perturbation. For simplicity, we will consider the hypersurface $t=t_0$ where $a(t_0)=1$, therefore $\dot{a}(t_0)=\kkappa=\sqrt{\ee_0/3}$.

  Inserting Eqns \eqref{eq:FLRWspin1}-\eqref{eq:FLRWspin2} into the constraints, \eqref{eq:phN}-\eqref{eq:phK} turn into
  \begin{align}
   2\frac{\partial_r\Nhat}{r}&=\frac{\Nhat^2(\eth\ethb\Nhat+\ethb\eth\Nhat)+2\Nhat+2(\kk\kkbar-1)\Nhat^3}{2r^2}-(\kkappa\KK+\tfrac{1}{4}\KK^2-\ee)\Nhat^3,\label{eq:HamFLRW}\\
   \partial_r\kk&=\Nhat(-\pp+\eth\kkappa+\tfrac{1}{2}\eth\KK)+(\kkappa-\tfrac{1}{2}\KK)\eth\Nhat-\frac{2\kk}{r},\label{eq:MomkFLRW}\\
   \partial_r\KK&=\Nhat\pph+\frac{2\kkappa-\KK}{r}+\frac{\mathrm{Re}[2\kkbar\eth\Nhat+\Nhat\eth\kkbar]}{r^2}.\label{eq:MomKFLRW}
  \end{align}
  Since $\Kstar>0$ for every $r>0$ we can integrate Eqns \eqref{eq:HamFLRW}-\eqref{eq:MomKFLRW} in the outward radial direction. Also the equations are formally singular at $r=0$. This is not merely a coordinate artifact, although not physical either, but related to the chosen foliation. Since the $r=const$ foliating surfaces degenerate into a point at the origin, the normal $1$-form is ill-defined there. In the rest of this section we will discuss how to handle this singularity.

  \subsection{Regularity at the origin}
  \label{sec:orig}

  In this section we present a procedure that determines the initial data, $(\Nhat|_{0},\kk|_{0},\allowbreak\KK|_{0})$, along with the derivatives, $(\partial_r\Nhat|_0,\partial_r\kk|_0,\partial_r\KK|_0)$, that ensures regularity of the equations at the origin. Somewhat surprisingly once we specify the free data, and hence the perturbations, the initial data $(\Nhat|_{0},\kk|_{0},\allowbreak\KK|_{0})$ and the derivatives $(\partial_r\Nhat|_0,\partial_r\kk|_0,\partial_r\KK|_0)$ are uniquely determined as the only solution compatible with regularity at the origin.

  We assume that the apparent singularity is entirely explained by the degeneracy of our foliation, and there is no physical singularity in the problem. Assuming that the fundamental variables are regular at the origin we derive the conditions that ensure the regularity of the derivatives $(\partial_r\Nhat|_{0},\partial_r\kk|_{0},\partial_r\KK|_{0})$ as limits of Eqns \eqref{eq:HamFLRW}-\eqref{eq:MomKFLRW}.

  We first consider Eqn \eqref{eq:MomkFLRW}. Taking the $r\rightarrow0$ limit yields
  \begin{equation}
   \partial_r\kk|_0-\Nhat|_0(-\pp|_0+\eth\kkappa|_0+\tfrac{1}{2}\eth\KK|_0)-(\kkappa|_0-\tfrac{1}{2}\KK|_0)\eth\Nhat|_0+\lim_{r\rightarrow0}\frac{2\kk}{r}=0.\label{eq:limkk1}
  \end{equation}
  The last term is indeterminate whenever $\kk|_0=0$. Then we apply l'H\^opital's rule $\lim_{r\rightarrow0}2\kk/r=2\partial_r\kk|_0$, and
  \begin{equation}
   \partial_r\kk|_0=\tfrac{1}{3}\left(\Nhat|_0(-\pp|_0+\eth\kkappa|_0+\tfrac{1}{2}\eth\KK|_0)+(\kkappa|_0-\tfrac{1}{2}\KK|_0)\eth\Nhat|_0\right).\label{eq:limkk2}
  \end{equation}
  We will further simplify this expression once we derive regularity conditions on $\Nhat$ and $\KK$.

  Next, we multiply both sides of Eqn \eqref{eq:HamFLRW} by $r$. After rearranging and taking the limit
  \begin{equation}
      \partial_r\Nhat|_0-\lim_{r\rightarrow0}\frac{\Nhat^2(\eth\ethb\Nhat+\ethb\eth\Nhat)+2\Nhat+2(\kk\kkbar-1)\Nhat^3}{4r}=0.
    \label{eq:limNh1}
  \end{equation}
  Using $\kk|_0=0$ the last term is of indeterminate form when
  \begin{equation}
    \Nhat|_0^2(\eth\ethb\Nhat|_0+\ethb\eth\Nhat|_0)+2\Nhat|_0-2\Nhat|_0^3=0
    \label{eq:limNh2}
  \end{equation}
  holds. Inserting the unit sphere Laplacian, which can be written as $\Delta=\frac{1}{2}(\eth\ethb+\ethb\eth)$ when acting on spin-weight $0$ quantities, we obtain
  \begin{equation}
   2\Nhat|_0^2\Delta\Nhat|_0+2(\Nhat|_0-\Nhat|_0^3)=0.
   \label{eq:limNh3}
  \end{equation}
  In spherical symmetry, when $\eth\Nhat=0$, the solutions are $\Nhat|_0=0,\pm1$. Among them we choose $\Nhat|_0=1$, which corresponds to a flow in the increasing $r$ direction. Using a maximum principle argument we can also show that Eqn \eqref{eq:limNh3} has no other solutions: Assuming that $\Nhat|_0$ is positive everywhere, we can divide by $\Nhat|_0$ and arrive at
  \begin{equation}
   -\Nhat|_0\Delta\Nhat|_0=1-\Nhat|_0^2.
  \end{equation}
  If $\Nhat|_0$ attains its maximum at point $p$, then $\Delta\Nhat|_{0,p}\leq0$ and $1\geq\Nhat|_{0,p}$ follows. Similarly if $\Nhat|_0$ has a minimum at point $q$, then $1\leq\Nhat|_{0,q}$. Therefore, $\Nhat|_0=1$ must hold. Then applying l'H\^opital's rule, and using $\kk|_0=0$ and $\Nhat|_0=1$ Eqn \eqref{eq:limNh1} gives
  \begin{equation}
      \partial_r\Nhat|_0-\tfrac{1}{4}\Delta\partial_r\Nhat|_0=0,
    \label{eq:limNh4}
  \end{equation}
  with the only solution $\partial_r\Nhat|_0=0$. Now Eqn \eqref{eq:limkk2} simplifies to
  \begin{equation}
      \partial_r\kk|_0=\tfrac{1}{3}(-\pp|_0+\eth\kkappa|_0+\tfrac{1}{2}\eth\KK|_0).
    \label{eq:limkk3}
  \end{equation}

  Determining $\KK|_0$ and $\partial_r\KK|_0$ is more complicated, because it is intertwined with $\partial_r\kk|_0$ and $\partial_r{}^2\kk|_0$. We start by multiplying both sides of Eqn \eqref{eq:MomKFLRW} by $r$ and taking the limit to get
  \begin{equation}
      2\kkappa|_0-\KK|_0+\lim_{r\rightarrow0}\frac{\mathrm{Re}[2\kkbar\eth\Nhat+\Nhat\eth\kkbar]}{r}=0.
    \label{eq:limKK1}
  \end{equation}
  Due to $\eth\Nhat|_0=0$ and $\kk|_0=0$ the last term is of indeterminate form and the limit gives
  \begin{equation}
      2\kkappa|_0-\KK|_0+\tfrac{1}{2}(\eth\partial_r\kkbar|_0+\ethb\partial_r\kk|_0)=0,
    \label{eq:limKK2}
  \end{equation}
  and
  \begin{equation}
      \KK|_0=2\kkappa|_0+\tfrac{1}{2}(\eth\partial_r\kkbar|_0+\ethb\partial_r\kk|_0).
    \label{eq:limKK3}
  \end{equation}

  Inserting Eqn \eqref{eq:limKK3} into Eqn \eqref{eq:limkk3} yields
  \begin{equation}
      \partial_r\kk|_0-\tfrac{1}{12}(\eth\eth\partial_r\kkbar|_0+\eth\ethb\partial_r\kk|_0)=\tfrac{1}{3}(-\pp|_0+2\eth\kkappa|_0).
    \label{eq:limkk4}
  \end{equation}
  This is a linear equation for $\partial_r\kk|_0$ and can be solved by inserting a spherical harmonic expansion $\partial_r\kk|_0=\sum [\partial_r\kk|_0]_\ell{}^m\cdot{}_1Y_\ell{}^m$. Then for each $(\ell,m)$ mode we get
  \begin{equation}
      [\partial_r\kk|_0]_\ell{}^m=\frac{\big[-\eth\eth\overline{\pp}|_0+2\eth\ethb\eth\overline{\kkappa}|_0+\eth\ethb\pp|_0-12\pp|_0-2\eth\ethb\eth\kkappa|_0+24\eth\kkappa|_0\big]_\ell{}^m}{6(6+\ell+\ell^2)}.
    \label{eq:limkk5}
  \end{equation}
  Since here $\kkappa$ takes the FLRW value its derivative is vanishing. Later, in Sec \ref{sec:num} we only consider $\pp=0$ perturbations. Then $\partial_r\kk|_0=0$ and $\KK|_0=2\kkappa|_0$.

  Now taking the limit of Eqn \eqref{eq:MomKFLRW}
  \begin{equation}
      \partial_r\KK|_0=\Nhat|_0\pph|_0+\lim_{r\rightarrow0}\left(2\kkappa-\KK+\frac{\mathrm{Re}[2\kkbar\eth\Nhat+\Nhat\eth\kkbar]}{r}\right)\bigg/r
    \label{eq:limKK4}
  \end{equation}
  We just ensured that the last term is of indeterminate form and
  \begin{equation}
      2\partial_r\KK|_0=2\partial_r\kkappa|_0+\pph|_0+\lim_{r\rightarrow0}\partial_r\frac{\mathrm{Re}[2\kkbar\eth\Nhat+\Nhat\eth\kkbar]}{r}
    \label{eq:limKK5}
  \end{equation}
  follows. It is straightforward to show that whenever $B|_0=0$ and $\partial_rB|_0$ is finite then $\lim_{r\rightarrow0}\partial_rB/r$ takes the value $\partial_{r}{}^2B|_0/2$. Thus
  \begin{equation}
      \partial_r\KK|_0=\partial_r\kkappa|_0+\tfrac{1}{2}\pph|_0+\tfrac{1}{8}(\eth\partial_r{}^2\kkbar|_0+\ethb\partial_r{}^2\kk|_0).
    \label{eq:limKK6}
  \end{equation}
  We can deduce $\partial_r{}^2\kk|_0$ by taking the derivative of Eqn \eqref{eq:MomkFLRW} and evaluating the limit. For the sake of simplicity we again restrict ourselves to the assumptions we will use in Sec \ref{sec:num}: $\pp=0$, $\kkappa=const$, and $\pph|_0=0$. Then $\partial_r{}^2\kk|_0=\eth\partial_r\KK|_0/4$, and $\partial_r\KK|_0=\Delta\partial_r\KK|_0/16$ which has the solution $\partial_r\KK|_0=0$.

  In summary with the assumptions $\kkappa=const$, $\pp=0$, and $\pph|_0=0$ we have
  \begin{align}
   \Nhat|_0&=1,&\partial_r\Nhat|_0&=0,\label{eq:regularity1}\\
   \kk|_0&=0,&\partial_r\kk|_0&=0,\label{eq:regularity2}\\
   \KK|_0&=2\kkappa|_0=2\sqrt{\ee_0/3},&\partial_r\KK|_0&=0.\label{eq:regularity3}
  \end{align}
  Although conditions \eqref{eq:regularity1}-\eqref{eq:regularity3} specify the initial data and the value the equations take at the origin, we still have to talk about regularity of the perturbations $\ee$ and $\pph$ subject to standard regularity conditions at the origin. Slightly generalizing the argument of \cite{boydChebyshevFourierSpectral1989} we deduce the following properties. As $\ee$ is a smooth scalar field the expansion coefficients $\ee_\ell{}^m(r)$ must have order $\ell$ zeros at the origin. Furthermore, $\ee_\ell{}^m(r)$ is symmetric to the origin if $\ell$ is even, and antisymmetric if $\ell$ is odd. Therefore only $\ee_0{}^0|_0$ has nonzero value and we have to take parity properties of the coefficients into account when thinking about the limits of derivatives. As opposed to this $\pph$ is a radial component of a $3$-vector\footnote{Strictly speaking the radial component is $r^aS_a$, but using FLRW free data $\pph\sim r^aS_a$.}. Its $\ell=0$ mode has a leading order $r$ behavior, while its $\ell=1$ mode might have a nonzero limit and the same parity argument applies. This justifies the $\pph|_0=0$ assumption we used above.

  The process presented in this section is equivalent to substituting a power series expansion of $(\Nhat,\kk,\KK)$, grouping the terms by powers of $r$ and solving the equations order by order as in \cite{Csukas:2025eer}. Since this latter method is easier to turn into an algorithm we discuss its application in the appendix.

  \section{Numerical studies}
  \label{sec:num}

  We apply the ideas presented in the previous section to numerically construct cosmological initial data, where anisotropy is introduced through axially symmetric perturbations in the matter variables $\ee$ and $\pph$.

  \subsection{Numerical methods}

  In our numerical studies we utilize the \texttt{ConstraintSolver} code \cite{CsukasKarolyZoltan2024}. All $f(\vartheta,\varphi)$ functions on the $\mathcal{S}_r$ spheres are represented as a truncated series
  \begin{equation}
   f(\vartheta,\varphi)=\sum_{\ell=|s|}^{\ell_{max}}\sum_{m=-\ell}^{\ell}f_\ell{}^m\cdot {}_sY_\ell{}^m(\vartheta,\varphi),
  \end{equation}
  where ${}_sY_\ell{}^m(\vartheta,\varphi)$ is the spin-weight $s$ spherical harmonic and products are re-expanded using Gaunt coefficients. Due to the use of spectral methods we expect spectral convergence in truncation order $\ell_{max}$.

  Ordinarily \texttt{ConstraintSolver} uses Runge--Kutta methods for integrating the constraints. However, Eqns \eqref{eq:HamFLRW}-\eqref{eq:MomKFLRW} are not smooth enough at the origin and the 4th order Runge--Kutta method loses an order of convergence. To avoid this loss of accuracy the first step from the origin is computed using Taylor series expansion around the origin. The expansion coefficients are computed by repeating the process discussed in Sec \ref{sec:orig} for the radial derivatives of the equations. The derivation of expansion coefficients is discussed in the appendix. After this first step evaluated by series expansion we expect 4th order algebraic convergence in $\Delta r$ step size.

  \subsection{Structure of perturbations}

  Our main objective is to insert localized inhomogeneities in the energy density and flux, and solve the constraints for the perturbed variables $(\Nhat,\kk,\KK)$. Here our choices are motivated by the combination of finite-difference and spectral methods.

\begin{figure}[H]
   \centering
   \begin{subfigure}{0.45\textwidth}
    \includegraphics[width=\textwidth]{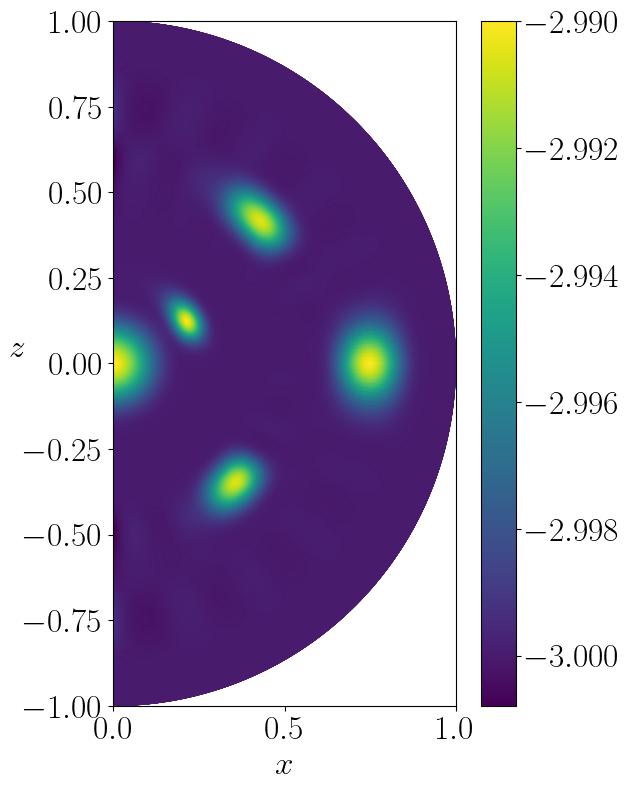}
    \caption{$\delta\ee$}
    \label{fig:ee}
   \end{subfigure}\hfill
   \begin{subfigure}{0.49\textwidth}
    \includegraphics[width=\textwidth]{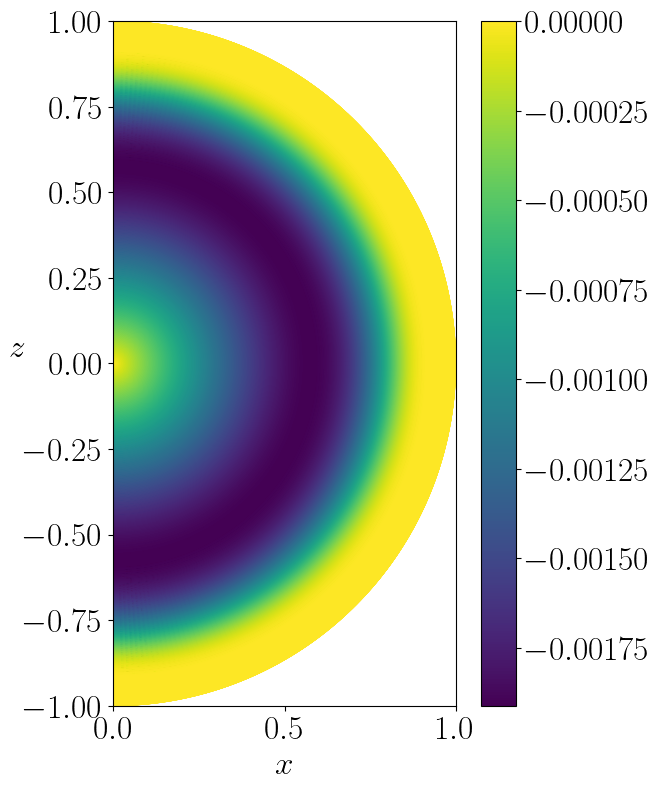}
    \caption{$\pph$}
    \label{fig:pr}
   \end{subfigure}
   \caption{Perturbation in the $\varphi=0$ plane. Panel \ref{fig:ee} shows the energy density perturbation $\delta\ee$. The perturbation consists of $5$ well-localized rings---although the effects of spectral truncation are still visible. Panel \ref{fig:pr} shows the energy flux perturbation $\pph$. Since the purpose of this perturbation is merely to induce some non-trivial structure in $\KK$, $\pph$ is just a simple spherically symmetric perturbation.}
   \label{fig:pert}
  \end{figure}

  In the radial direction the perturbations are represented by the bump function
  \begin{equation}
   \mathcal{B}(r)=
   \begin{cases}
    \exp\left(-\frac{1}{r-c+w/2}+\frac{1}{r-c-w/2}+4/w\right)&c-w/2\leq r\leq c+w/2\\
    0&\text{otherwise}
   \end{cases}.
  \end{equation}
  However, in the angular direction we use a fixed $\ell_{max}=20$ truncated representation of the Gaussian
  \begin{equation}
   \mathcal{G}(\vartheta)=\exp\left(-\frac{(\cos\vartheta-c)^2}{2w^2}\right).
  \end{equation}
  For the sake of simplicity we only consider perturbations that are negligible on the poles. The truncation error of $\mathcal{G}$ falls off faster than that of a $\mathcal{B}$ bump function, but the effect of truncation is still visible at $\ell_{max}=20$. We use a $5$ term perturbation, where each term is a product of a bump and a Gaussian, with an overall scale factor. Specifically we take
  \begin{equation}
   \delta\ee=10^{-2}\left(\sum_{n=0}^4\mathcal{B}_{n}(r)\cdot\mathcal{G}_{n}(\vartheta)\right),
  \end{equation}
  where the center and width of the bumps and Gaussians are displayed in Tab \ref{tab:cwBumpGauss}. The structure of the resulting perturbation $\delta\ee$ is depicted in Fig \ref{fig:ee}.

  \begin{table}
  \begin{center}
   \begin{tabular}{c|c|c|c|c}
    $n$&$c_{\mathcal{B}}$&$w_{\mathcal{B}}$&$c_{\mathcal{G}}$&$w_{\mathcal{G}}$\\\hline
    $0$&$0$&$0.6$&-&-\\
    $1$&$0.25$&$0.3$&$0.5$&$0.15$\\
    $2$&$0.5$&$0.4$&$-0.7$&$0.08$\\
    $3$&$0.75$&$0.5$&$0$&$0.1$\\
    $4$&$0.6$&$0.4$&$0.7$&$0.08$
   \end{tabular}
  \end{center}
  \caption{Center and width of the applied bumps and Gaussians in $\delta\ee$. The missing parameters in the first row means that we take $\mathcal{G}_0=1$.}
  \label{tab:cwBumpGauss}
  \end{table}

  Notice that in the absence of further perturbations the initial data is exactly the background value. Furthermore notice that the terms on the right hand side of Eqns \eqref{eq:MomkFLRW}-\eqref{eq:MomKFLRW} are trivial when they take the background value. It follows that in the absence of further perturbations only $\Nhat$ will display any difference from the background. To produce a more interesting solution we also introduce a simple
  \begin{equation}
   \pph=10^{-3}\partial_r\mathcal{B}
  \end{equation}
  perturbation with center $c=0$ and width $w=2$, whose sole purpose is to drive $\KK$ away from the background value and turn on non-trivial perturbations in $\KK$ and $\kk$. This perturbation in $\pph$ is depicted in Fig \ref{fig:pr}.

  \subsection{Results}

  Although the Gaussian is truncated at $\ell=20$ the actual computations use $\ell_{max}=22$ as base resolution, so we have at least two ``empty'' modes. This is to lessen the effect of the truncation error of the perturbations. The Taylor step and the Runge--Kutta steps are taken with $\Delta r=10^{-4}$ for the reference data. Setting $\ee_0=3$, we integrate until the Hubble radius of the background, which in the chosen geometric units, corresponds to $r=1$. The regularity conditions together with the perturbations specified above and the value of $\ee_0$ result in the initial data $\Nhat|_0=1$, $\KK|_0=2$, and $\kk|_0=0$.

  \begin{figure}[H]
   \centering
   \begin{subfigure}{0.3\textwidth}
    \includegraphics[width=\textwidth]{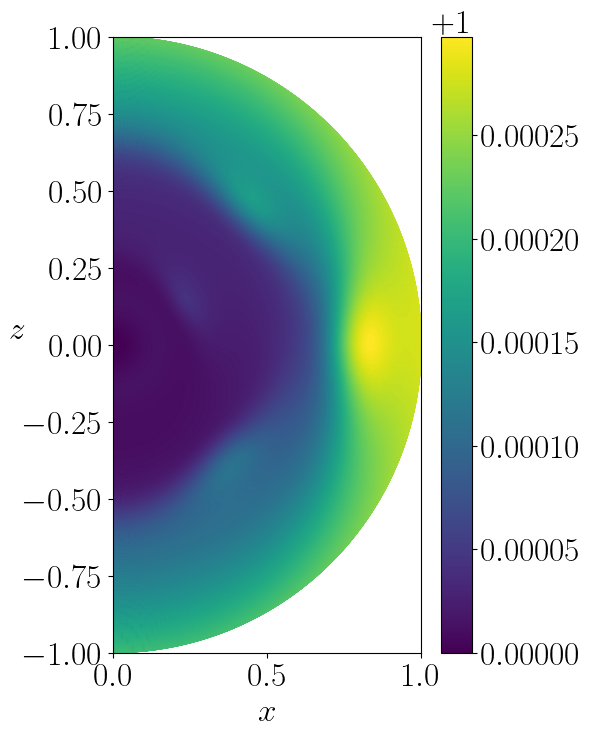}
    \caption{$\Nhat$}
    \label{fig:Nhat}
   \end{subfigure}\hfill
   \begin{subfigure}{0.3\textwidth}
    \includegraphics[width=\textwidth]{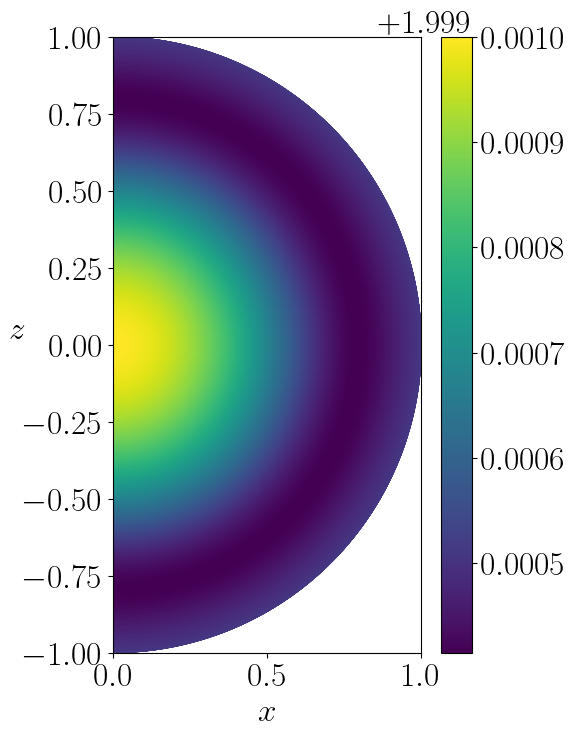}
    \caption{$\KK$}
    \label{fig:KK}
   \end{subfigure}\hfill
   \begin{subfigure}{0.3\textwidth}
    \includegraphics[width=\textwidth]{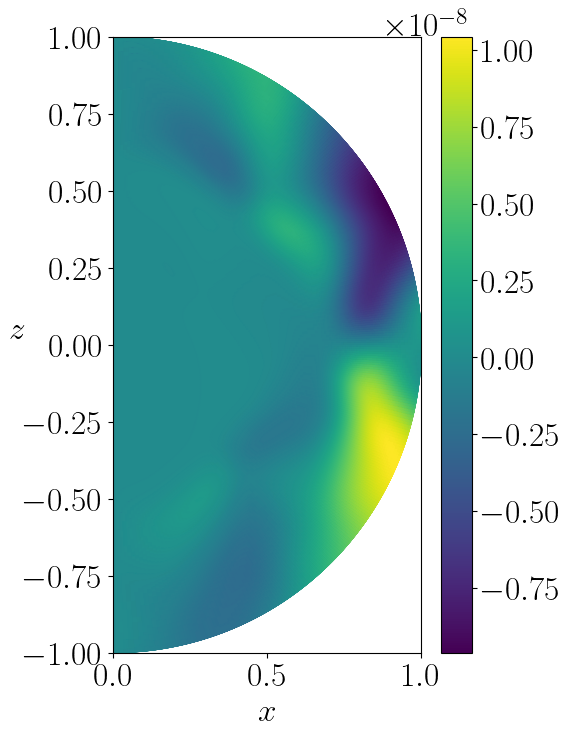}
    \caption{$\mathrm{Re}[\kk]$}
    \label{fig:kk}
   \end{subfigure}
   \caption{The constrained variables in the $\varphi=0$ plane. Panel \ref{fig:Nhat} shows the perturbed $\Nhat$. Due to the perturbation the variable slowly increases, but the increase is not isotropic. We can clearly see the trace of the perturbation. Panel \ref{fig:KK} shows the perturbed $\KK$. The effect of the spherically symmetric perturbation $\delta\pph$ drives $\KK$ away from the homogeneous background. Unlike $\Nhat$ here we see that after reaching a maximum $\KK$ starts decreasing again. Panel \ref{fig:kk} shows the perturbed $\kk$. Due to the symmetries of the problem only the real part of $\kk$ attains nonzero value. Although we did not perturb $\kk$ directly, the most interesting structure can be seen here. This is related to $\kk$ being a spin-weight $1$ variable.}
   \label{fig:soln}
  \end{figure}

  Fig \ref{fig:soln} shows the constrained variables generated by the perturbations on the $\varphi=0$ section. Fig \ref{fig:Nhat} shows that the perturbation introduced by $\delta\ee$ induced an increase in $\Nhat$ as we get farther from the origin, with the contour lines clearly deformed according to the structure of the perturbation. Similarly in Fig \ref{fig:KK} the perturbation introduced by $\pph$ drives $\KK$ away from the background value but the change is less monotonic than in the previous case. The most interesting behavior is depicted in Fig \ref{fig:kk}. Although $\kk$ remains small, it displays an unexpectedly rich structure.

  \begin{figure}[H]
   \centering
   \begin{subfigure}{0.45\textwidth}
    \includegraphics[width=\textwidth]{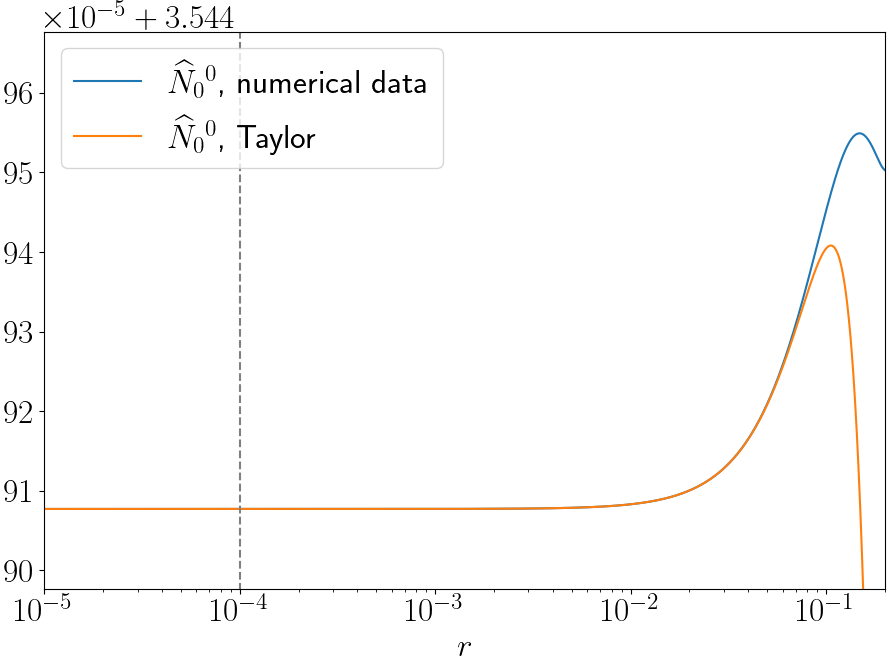}
    \caption{$\Nhat_0{}^0$}
    \label{fig:Nh00}
   \end{subfigure}\hfill
   \begin{subfigure}{0.45\textwidth}
    \includegraphics[width=\textwidth]{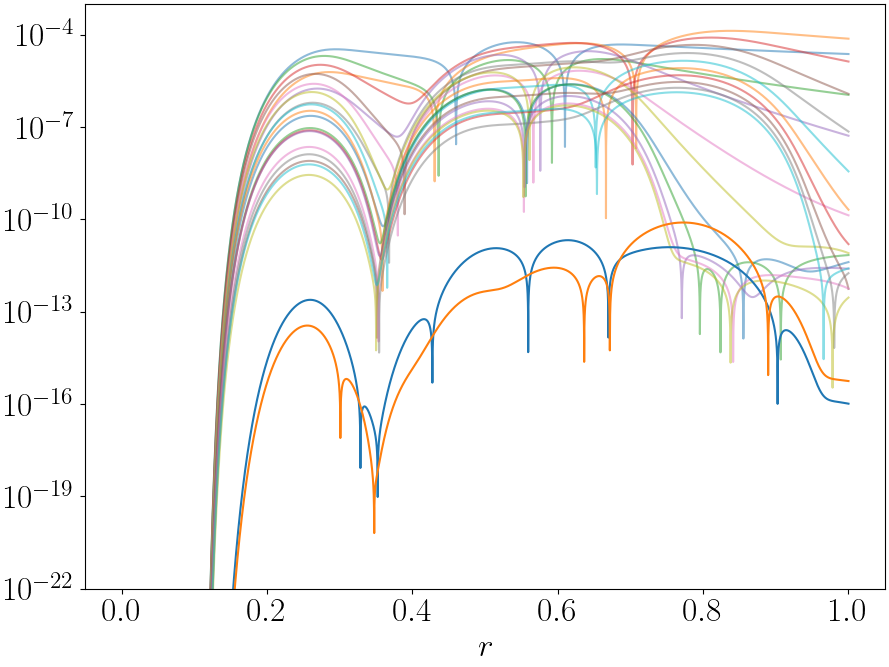}
    \caption{$\Nhat_\ell{}^0$}
    \label{fig:Nhmode}
   \end{subfigure}\\
   \begin{subfigure}{0.45\textwidth}
    \includegraphics[width=\textwidth]{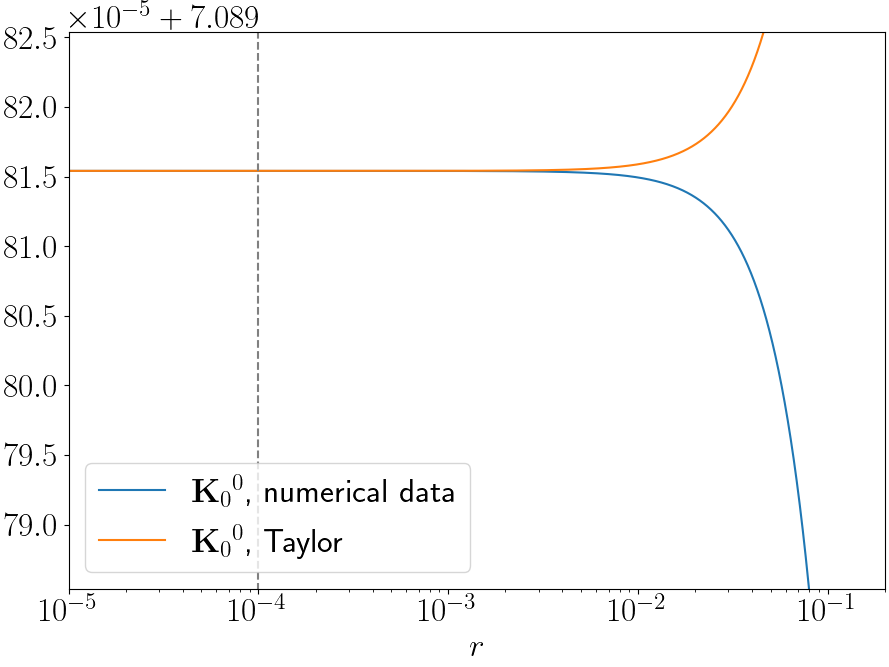}
    \caption{$\KK_0{}^0$}
    \label{fig:KK00}
   \end{subfigure}\hfill
   \begin{subfigure}{0.45\textwidth}
    \includegraphics[width=\textwidth]{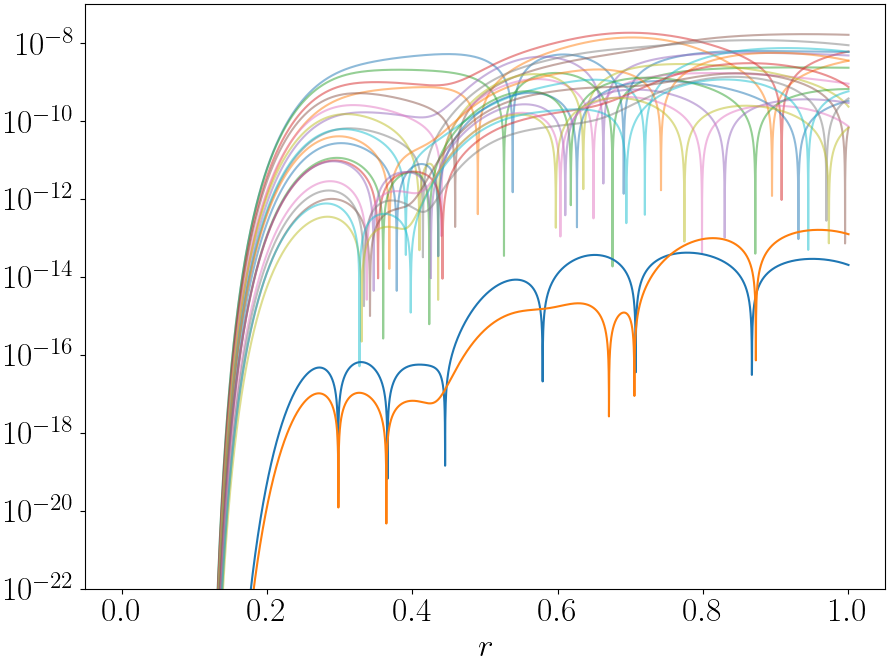}
    \caption{$\KK_\ell{}^0$}
    \label{fig:KKmode}
   \end{subfigure}\\
   \begin{subfigure}{0.45\textwidth}
    \includegraphics[width=\textwidth]{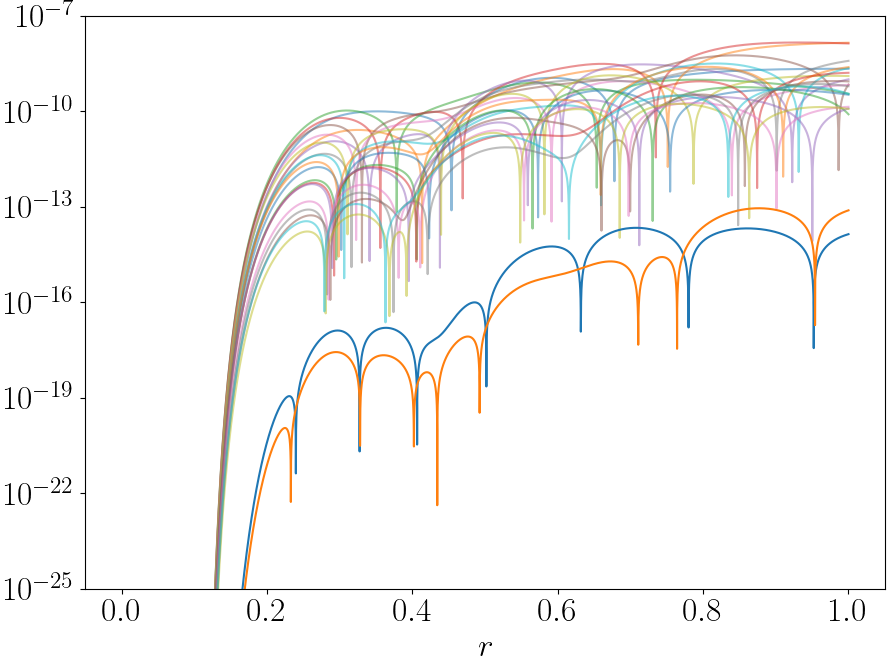}
    \caption{$\mathrm{Re}[\kk_\ell{}^0]$}
    \label{fig:kkmode}
   \end{subfigure}
   \caption{Panels \ref{fig:Nh00} and \ref{fig:KK00} show that the applied Taylor series expansion agrees well with the numerical data to significantly larger $r$ than where it was applied (shown by dashed vertical lines). Panels \ref{fig:Nhmode}, \ref{fig:KKmode}, and \ref{fig:kkmode} show the non-vanishing $\ell>0$ modes of $\Nhat$, $\KK$, and $\kk$. First, we see why we had to use $\ell_{max}=22$ with perturbation truncated at $\ell=20$. The modes with $\ell\leq20$, shown by pale colors, are excited by the perturbation which yields a relatively large truncation error. As opposed to this, the $\ell=\{21,22\}$ modes are significantly smaller, letting us observe a clearer angular convergence. Second, the $\ell>0$ modes of  $\Nhat$ fall off after we leave behind the perturbation. This suggests that the solution might asymptote to an FLRW solution. Since $\pph$ is non-vanishing throughout the domain we do not observe similar fall off behavior in $\KK$ and $\kk$.}
   \label{fig:modes}
  \end{figure}

  A more informative mode-by-mode depiction can be seen in Fig \ref{fig:modes}. Panels \ref{fig:Nh00} and \ref{fig:KK00} demonstrate that the Taylor series expansion derived in the appendix provides a good approximation of the solution near the origin. The agreement is convincing well past $r=10^{-4}$, where it was applied. Higher $\ell$ modes of $\Nhat$, $\KK$, and $\kk$ are shown in Panels \ref{fig:Nhmode}, \ref{fig:KKmode}, and \ref{fig:kkmode} respectively. We see that $\ell>0$ modes in $\Nhat$ display a fall off behavior for values larger than $r\approx0.8$, once we leave the region of the effective perturbations behind. This suggests that the solution might asymptote to an FLRW solution. Similar behavior cannot be observed in $\KK$ and $\kk$, probably because $\pph$ extends to the whole domain. The question of the asymptotic behaviour of solutions is left for a future study.

  \begin{figure}[H]
   \centering
   \begin{subfigure}{0.5\textwidth}
    \includegraphics[width=\textwidth]{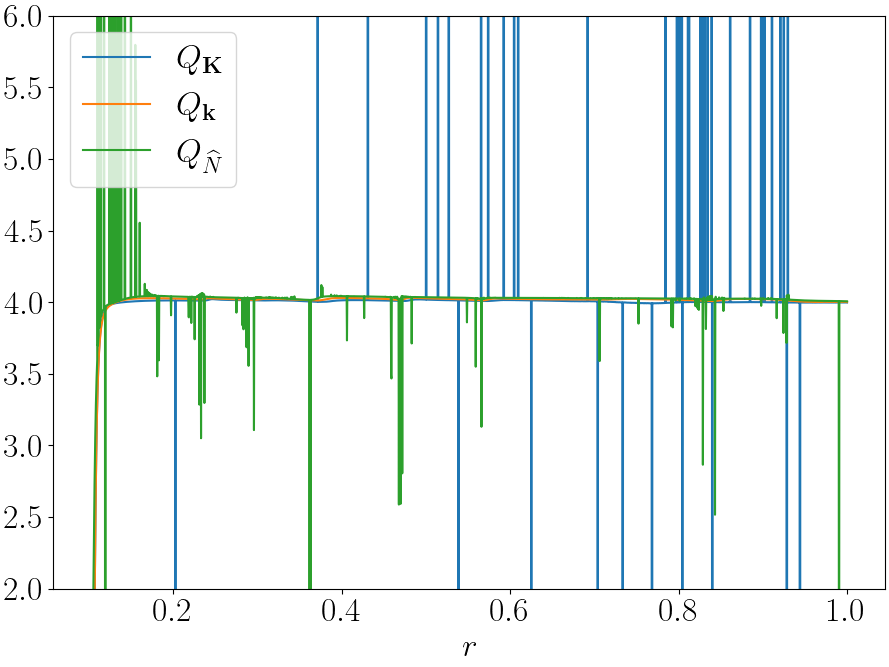}
    \caption{Convergence in radial resolution.}
    \label{fig:rconv}
   \end{subfigure}\hfill
   \begin{subfigure}{0.5\textwidth}
    \includegraphics[width=\textwidth]{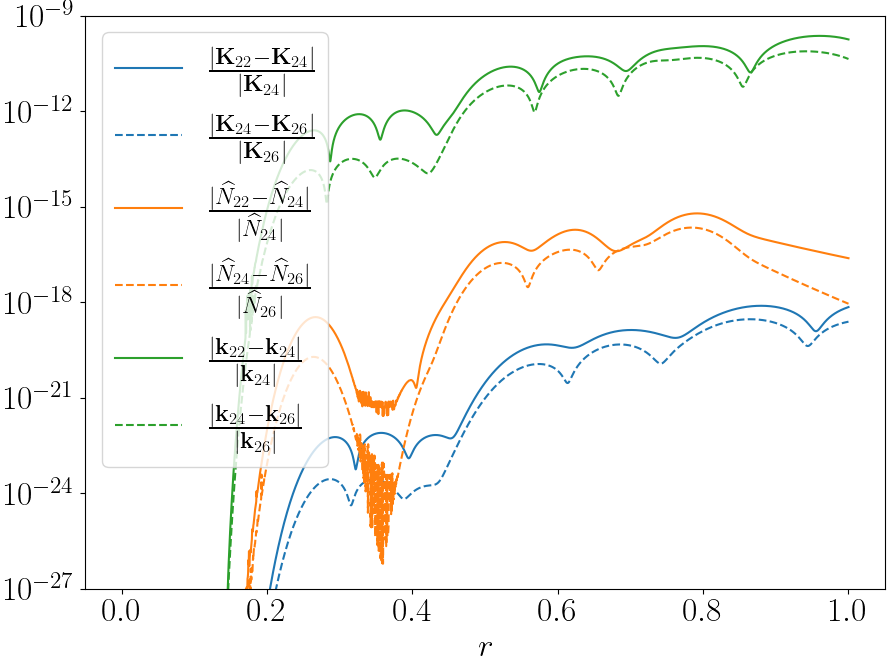}
    \caption{Convergence in angular resolution.}
    \label{fig:lconv}
   \end{subfigure}
   \caption{Convergence properties. Panel \ref{fig:rconv} shows the local convergence rate as we vary the step size in $r$. Apart from spikes due to roundoff errors we observe the expected 4th order algebraic convergence rate. Panel \ref{fig:lconv} shows error estimates as we change the angular resolution. The consistent shift in the error is the hallmark of spectral convergence.}
   \label{fig:conv}
  \end{figure}

  The convergence properties of the solution are shown in Fig \ref{fig:conv}. In the radial direction we observe the local rate of convergence by computing
  \begin{equation}
   Q_f=\log_2\frac{|f_4-f_2|_{L^2}}{|f_2-f_1|_{L^2}},
  \end{equation}
  where $f_1$ is data on the base resolution $\Delta r=10^{-4}$, $f_2$ is data computed using double the step size of the base resolution, $\Delta r=2\cdot10^{-4}$, and $f_4$ is computed using quadruple of the base resolution step size, $\Delta r=4\cdot10^{-4}$. $Q$ is evaluated on the coarsest grid with step size $\Delta r=4\cdot10^{-4}$. Fig \ref{fig:rconv} shows $Q$ for each constrained variable. Early on we do not see a convergence rate, because near the origin the discretization error already reaches machine precision with way larger $\Delta r=10^{-3}$ step size. When we reach the second bump function the solution becomes more complicated and the discretization error grows large enough to observe the expected algebraic convergence rate characteristic of the 4th order Runge--Kutta algorithm.

  Angular convergence is measured by comparing the truncation error, which is estimated by the $L^2$ norm of the differences between different resolution data normalized by the norm of the finer resolution data. Specifically we constructed solutions with fixed $\Delta r=4\cdot 10^{-4}$ and $\ell_{max}=\{22,24,26\}$. For a meaningful comparison we only consider spherical harmonic modes up to $\ell=22$ in each data set. In Fig \ref{fig:lconv} we observe the shift in the truncation error typical for spectral methods.

  \section{Discussion}
  \label{sec:discussion}

  In this paper we demonstrated the viability of a new method to construct cosmological initial data numerically. By solving the parabolic-hyperbolic form of the Einstein constraint equations we start the integration from the origin and generate the solution until a predetermined radius, which here corresponds to the Hubble-scale of the background. The integration can subsequently be continued either without or with additional compactly supported perturbations.

  The most notable features of the presented method stem from the combination of the parabolic-hyperbolic formulation and choosing a spherical foliation. This combination allows us to put aside the somewhat vaguely motivated boundary conditions of the conformal methods with well-motivated regularity conditions at the origin. Once we specify the desired perturbations in the matter sector both the field values and the Taylor series expansion of the solutions are uniquely determined in the whole domain. We also remind the reader that using evolutionary methods we always select a unique solution, unlike the conformal or elliptic methods that sometimes fail to find the solution because the lack of its uniqueness.

  Although in this paper we considered only a spatially flat background, the method generalizes readily to a spatially hyperbolic one. The crucial part of the analysis is determining the regularity conditions at the origin, which immediately applies to the hyperbolic case too. The $S^3$ topology, however, requires further investigation, as it has another distinguished location, the equator, which claims for a separate treatment. Generalization to $\Lambda\mathrm{CDM}$ is also straightforward. A spatially constant $\Lambda$ only changes the regularity condition as an additional contribution to $\ee_0$. Since we do not consider evolution of the initial data, where the effects of $\Lambda$ would show up, we can absorb it into $\ee_0$.

  Beyond providing an alternative method for constructing cosmological initial data, the present framework opens several directions for future work. Probably the most interesting application of the initial data generated by our method would be comparing its evolution with the evolution of initial data with various boundary conditions. There are methods, such as in \cite{Lim:2009dg,Coley:2023iol}, that can evolve the initial data without enforcing boundary conditions making such comparative studies feasible. Another application would be studying the formation of primordial black holes. Ref  \cite{Baumgarte:2025vvs} discusses non-uniqueness of the initial data in the context of primordial black holes and inflation, and they note that standard elliptic solvers may choose the wrong branch, from the perspective of the intended application. For example the solver chooses the branch tuned for black hole formation instead of inflation and vice versa. The method we introduced in this paper is free of these types of drawbacks and provide suitable initializations of various cosmological scenarios.

  \section*{Acknowledgments}

  The author thanks István Rácz and János Östör for the enlightening discussions and their feedback on the manuscript.

  This work was supported by the Hungarian Scientific Research Fund NKFIH Grant No. K-142423.

  \section*{AI statement}

  The author used OpenAI's GPT-5.6 Luna to check grammar and clarity, and identify typographical errors. No AI-assistance was used in generating code, data, and figures, and no AI tools were used in conceptualization or writing the first draft of the manuscript. The author reviewed and verified all AI-assisted output and takes full responsibility for the final manuscript.

  \section*{Software and data availability}

  The specific version of the software along with the necessary configuration files  and data will be available on Zenodo upon the acceptance of the paper.

  During the analysis the author made use of the following software. Figures were produced using \verb|matplotlib| \cite{Hunter:2007ouj}, \verb|numpy| \cite{Harris:2020xlr}, and \verb|pandas| \cite{teamPandasdevPandasPandas2025,McKinney:2010nts}. Some of the computations made use of \verb|Wolfram Mathematica| \cite{Mathematica}.

  \appendix

  \section{Taylor series around the origin}
  \label{sec:taylor}

  In this appendix we present some formulae used in the Taylor series expansion of the constrained variables used to take the first step away from $r=0$. The computation is analogous to the one presented in \cite{Csukas:2025eer}. We expand each constrained variable in a power series as
  \begin{align}
   \Nhat(r,\vartheta,\varphi)&=\sum_{n=0}^4\Nhat_n(\vartheta,\varphi)\cdot r^n+r^4\cdot w_{\Nhat}(r,\vartheta,\varphi),\\
   \kk(r,\vartheta,\varphi)&=\sum_{n=0}^4\kk_n(\vartheta,\varphi)\cdot r^n+r^4\cdot w_{\kk}(r,\vartheta,\varphi),\\
   \KK(r,\vartheta,\varphi)&=\sum_{n=0}^4\KK_n(\vartheta,\varphi)\cdot r^n+r^4\cdot w_{\KK}(r,\vartheta,\varphi),
  \end{align}
  and we insert them into Eqns \eqref{eq:HamFLRW}-\eqref{eq:MomKFLRW}. We also assume $\pp=0$ and $\pph$ being an odd function. After collecting by powers of $r$ we will have terms, whose $r$ dependence is explicit, and terms with implicit dependence, but of higher order. The lower order terms with explicit dependence must be equal on both sides of the equations. Here we discuss the resulting equations and their solutions.

  At leading order we get $\kk_0=0$,
  \begin{equation}
   \Nhat_0\left(\Nhat_0\Delta\Nhat_0-\Nhat_0^2+1\right)=0,
  \end{equation}
  with the positive solution $\Nhat_0=1$, and $\KK_0=2\kkappa_0+\frac{1}{2}(\eth\overline{\kk_1}+\ethb\kk_1)$ matching with Eqns \eqref{eq:limNh3} and \eqref{eq:limKK3}.

  Next to leading order we get
  \begin{align}
   -(\eth\eth\overline{\kk_1}+\eth\ethb\kk_1)+12\kk_1&=0,\\
   -\Delta\Nhat_1+4\Nhat_1&=0,
  \end{align}
  in accordance with Eqns \eqref{eq:limkk4} and \eqref{eq:limNh4}, which yields $\kk_1=0$, $\Nhat_1=0$, and simplifies the previous expression $\KK_0=2\kkappa_0$. We also get $\KK_1=\frac{1}{4}(\eth\overline{\kk_2}+\ethb\kk_2)$  in accordance with Eqn  \eqref{eq:limKK6}.

  Next we get
  \begin{align}
   -(\eth\eth\overline{\kk_2}+\eth\ethb\kk_2)+32\kk_2&=0,\\
   -\Delta\Nhat_2+6\Nhat_2&=\delta\ee_0,
  \end{align}
  which yields $\kk_2=0$, $\Nhat_{2,\ell,m}=\delta\ee_{0,\ell,m}/(\ell^2+\ell+6)$, and simplifies the previous expression $\KK_1=0$. We also get $\KK_2=\frac{1}{6}(\eth\overline{\kk_3}+\ethb\kk_3+2\pph_1)$.

  Next we get
  \begin{align}
   -(\eth\eth\overline{\kk_3}+\eth\ethb\kk_3)+60\kk_3-2\eth\pph_1&=0,\\
   -\Delta\Nhat_3+8\Nhat_3&=\delta\ee_1,
  \end{align}
  which yields
  \begin{align}
      \kk_{3,\ell,m}&=\frac{[\eth(\eth\ethb\overline{\pph_1}+60\pph_1-\eth\ethb\pph_1)]_\ell{}^m}{60 (30 + \ell + \ell^2)},\\
   \Nhat_{3,\ell,m}&=\frac{\delta\ee_{1,\ell,m}}{\ell^2+\ell+8},
  \end{align}
  and we also get $\KK_3=\frac{1}{8}(\eth\overline{\kk_4}+\ethb\kk_4)$.

  Finally
  \begin{equation}
   -(\eth\eth\overline{\kk_4}+\eth\ethb\kk_4)+96\kk_4=0
  \end{equation}
  yields $\kk_4=0$ simplifying $\KK_3=0$.

  \printbibliography
\end{document}